# Interoperable verification and dissemination of software assets in repositories using COAR Notify

*Matteo Cancellieri, CORE, The Open University, matteo.cancellieri@open.ac.uk ORCID: 0000-0002-9558-9772*

*Martin Docekal, Brno University of Technology, idocekal@fit.vutbr.cz, ORCID: 0000-0002-0580-9357*

*David Pride, CORE, KMi, The Open University david.pride@open.ac.uk ORCID: 0000-0002-7162-7252*

*Morane Gruenpeter, Software Heritage, morane@softwareheritage.org*

*David Douard, Software Heritage, david.douard@sdfa3.org*

*Petr Knoth, CORE, The Open University, petr.knoth@open.ac.uk ORCID: 0000-0003-1161-7359*

## Abstract

The discoverability, attribution, and reusability of open research software are often hindered by its obscurity within academic manuscripts. To address this, the SoFAIR project (2024–2025) introduces a comprehensive workflow leveraging machine learning tools for extracting software mentions from research papers. The project integrates repository systems, authors, and services like HAL and Software Heritage to ensure proper archiving, citation, and accessibility of research software in alignment with FAIR principles. To enable interoperable communication across the various systems we present an integration of the COAR Notify Protocol, which facilitates automated, interoperable communication among repositories and authors to validate and disseminate software mentions. This paper outlines the SoFAIR workflow and the implementation of the COAR Notify Protocol, emphasizing its potential to enhance the visibility and credibility of research software as first-class bibliographic records.

## Introduction

One of the primary challenges affecting the discoverability, attribution, and reusability of open research software lies in its frequent obscurity within research manuscripts. These valuable resources are often mentioned in passing or buried in supplementary material, preventing them from being recognized as distinct, citable outputs. For research software to achieve the status of first-class bibliographic records, it must first be systematically identified and assigned persistent identifiers (PIDs), ensuring alignment with FAIR (Findable, Accessible, Interoperable, and Reusable) principles. Despite their importance, many open research software assets still fall short of these principles, and explicit links between software resources and the papers introducing or utilizing them remain rare.

## Methodology

The SoFAIR project, spanning two years (2024-2025), aims to address this issue by leveraging the vast content available across the global network of open repositories.

In this paper, we will introduce a protocol to validate and disseminate the Software mentions discovered using the AI models on the project.

The SoFAIR workflow (Figure 1) illustrates the collaborative process between stakeholders, tools, and infrastructures to ensure research software is properly archived, cited, and made accessible. The workflow begins with an author depositing a piece of research software in a code repository (Step 1). Following this, the author submits a manuscript that includes either explicit or implicit references to the software (Step 2).

Subsequently, the research paper is harvested from the repository by CORE, where mentions of the software are extracted from the full text using advanced machine-learning tools such as GROBID and Softcite [1] (Step 3). These tools identify and structure software mentions for further processing.

Through the CORE Repository Dashboard[2], a request for validation of the extracted mentions is generated and made available to the repository (Step 4). With the repository manager's authorisation, this validation request is routed to the author, typically via an email notification, prompting them to validate the identified mentions (Step 5).

Upon successful validation, the repository issues an asset registration request to Software Heritage (Step 6). Software Heritage then permanently archives the software asset (Step 7), assigns a permanent identifier to it, and sends this identifier back to the repository (Step 8).

The whole workflow and details are documented in the SoFAIR project documentation page [8]

From the perspective of an aggregator, CORE possesses the capability to identify and extract software mentions at scale and it provides already a place to manage it in the CORE Dashboard[9] a service for repository managers to manage enrichments, check compliance and validate the quality of the repository metadata. This is achieved not only by cross-referencing duplicate instances of the same paper across multiple institutional repositories but also through the application of machine learning models specifically developed within the project.

From an aggregator's perspective, CORE can identify and extract software mentions at scale and it also provides a platform for managing these mentions through the CORE Dashboard [9], a service designed for repository managers to oversee enrichments, ensure compliance, and validate the quality of their repository metadata. The discovery of software mentions is enabled by combining cross-referencing duplicate instances of the same paper across multiple institutional repositories and employing machine learning models specifically developed within the project.

The discovery of software mentions does not inherently guarantee their accuracy or validity. To establish a reliable connection between a research paper and the mentioned software within the research graph, these mentions must be validated by an authoritative source, such as the paper's author or a designated research manager.

To facilitate the visibility and management of the discovered software mentions, we introduce the CORE Dashboard Software Mention tab. The tab (Figure 2 in the Appendix) displays the detected software mentions, offering repository managers the tools to process and integrate them either manually or through automated workflows.

Users can choose to automatically notify all authors of a paper by selecting a confidence threshold. The confidence score, determined by the AI tool employed, reflects the model's certainty regarding the accuracy of the software mentioned, with higher scores indicating greater confidence.

To enable an interoperable method of communication between the various components of the repository community we decided to integrate the communication mechanism using COAR Notify [3].

The COAR Notify Protocol establishes profiles, constraints, and conventions for utilizing W3C Linked Data Notifications (LDN)[4] to integrate repository systems with related services within a distributed, resilient, and web-native architecture.

The advantages of using COAR Notify come both from using a reusable standard already well-known in the repository community and also using a standard that has already a set of implementations in major repository software like DSpace[6] and is integrated already in major platforms like HAL[5].

When a notification is triggered, a COAR Notify message is sent to the Repository Inbox for processing. Within the repository, this processing facilitates the conversion between the paper and the approving actor. The implementation of this conversion may vary depending on the repository system but generally results in the generation of a message—delivered via email or an internal notification system—containing the details of the software mentions identified in the research paper.

The COAR Notify payload is inspired by the PCI Endorsement workflow pattern outlined by the COAR Notify use case [7] and it uses a simple pattern such as:

1. offer the relationship between a research paper and a software to validate
2. accept/tentatively accept/reject the relationship
3. announce the validated relationship

After the repository inbox processes the notification, it must support messaging to the actor responsible for validating the software mention, such as an author or a research manager either via email or through an internal communication system. We envisage the email content to be similar to the one described in Figure 3 in the Appendix.

The message will allow the possibility to:

- reject, in case it was incorrectly discovered
- edit, allowing the authors to add additional details to improve the software description.
- validate, storing the mention in the research graph.

Once the request is validated the information is stored by CORE and the software mention is announced and disseminated to anyone involved that subscribed to the software mentions.

The presentation will demonstrate and introduce the new pattern in COAR Notify and provide a description of the first full workflow aiming to use the project partners of HAL and their already implemented COAR-Notify Inbox [4] as a repository service and Software Heritage as an entity receiving the announcements of software mentions.

## Appendix: Figures and Tables

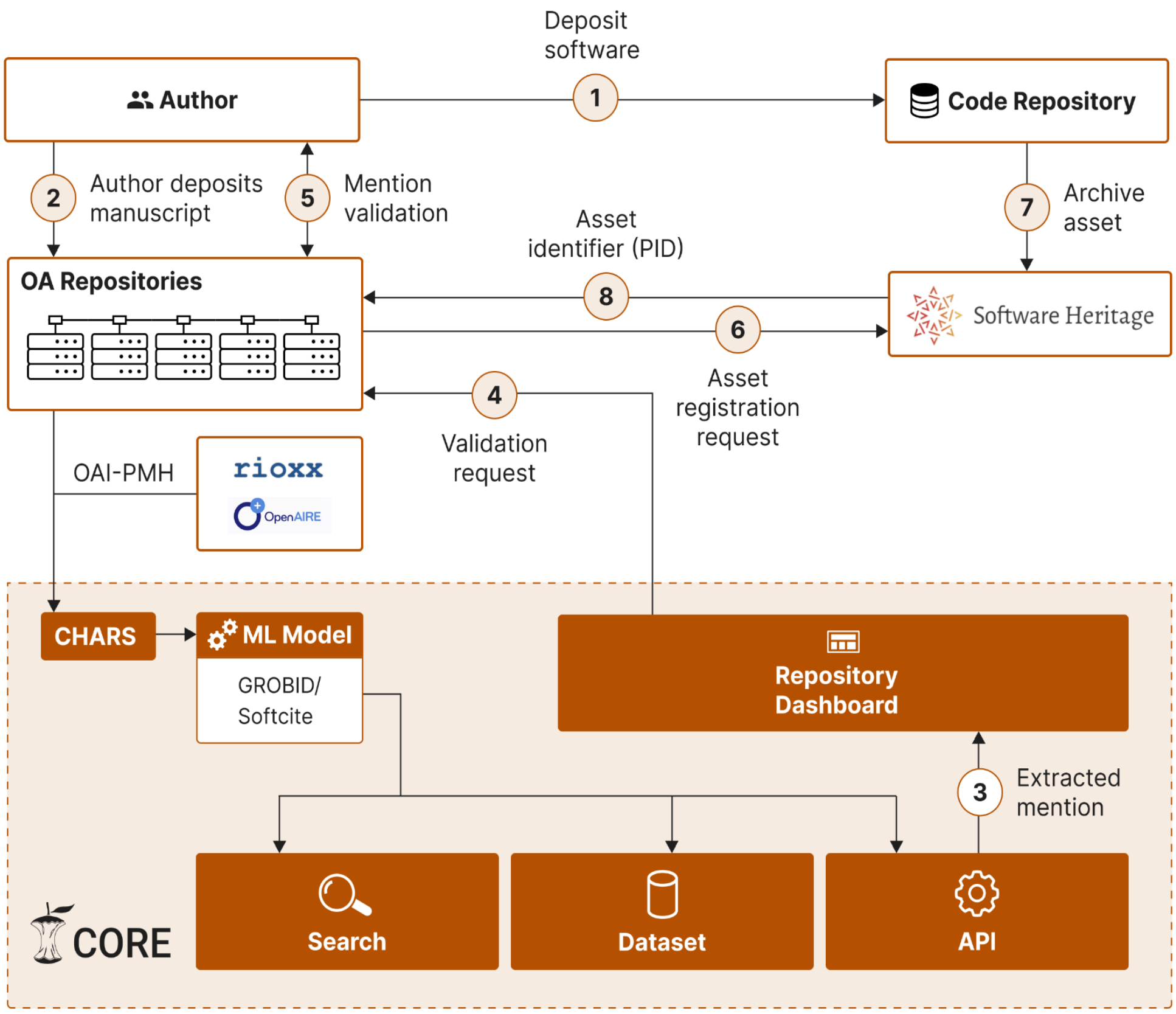

*Figure 1: The SoFAIR workflow*

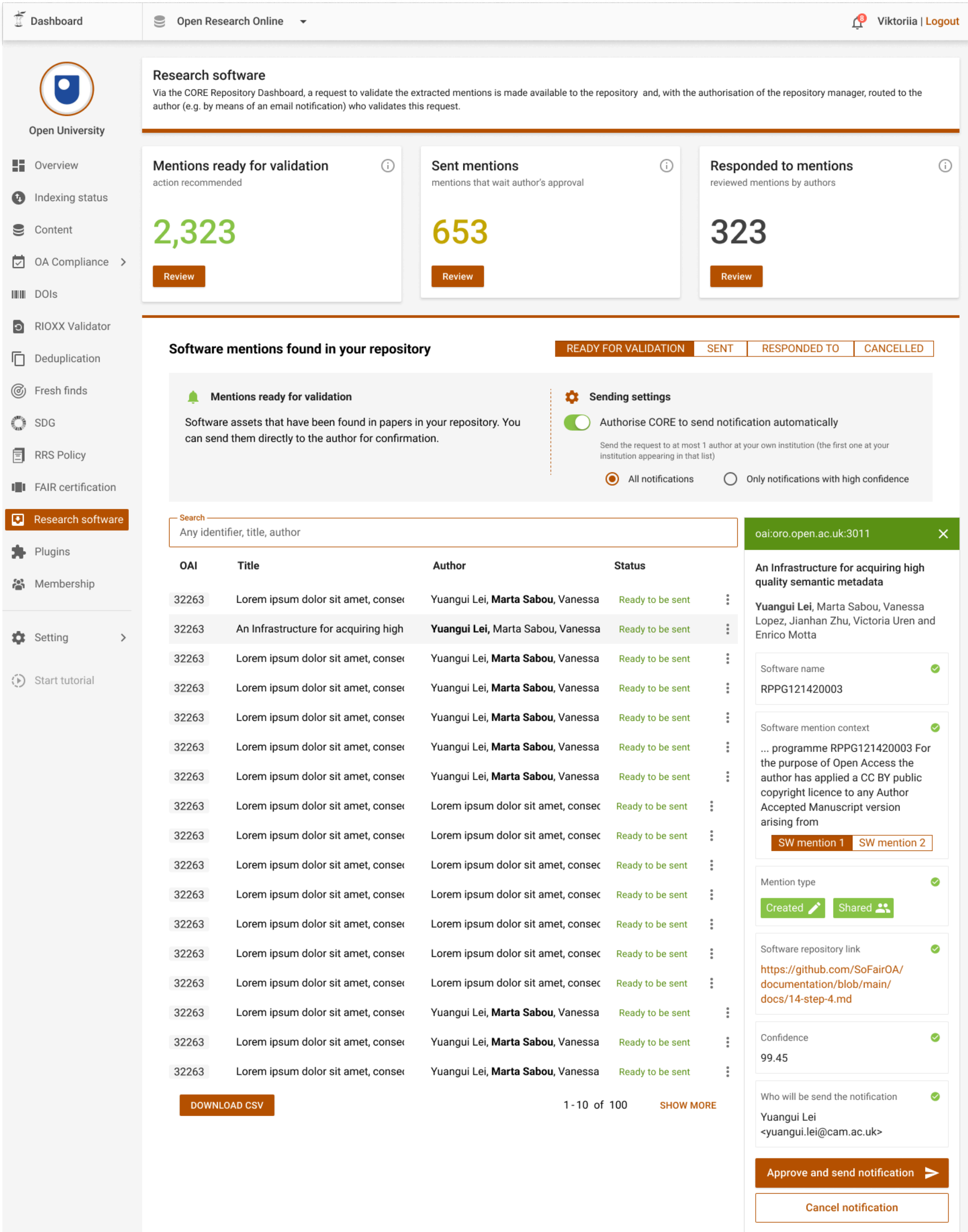

*Figure 2: The CORE Dashboard Software Mention tab*

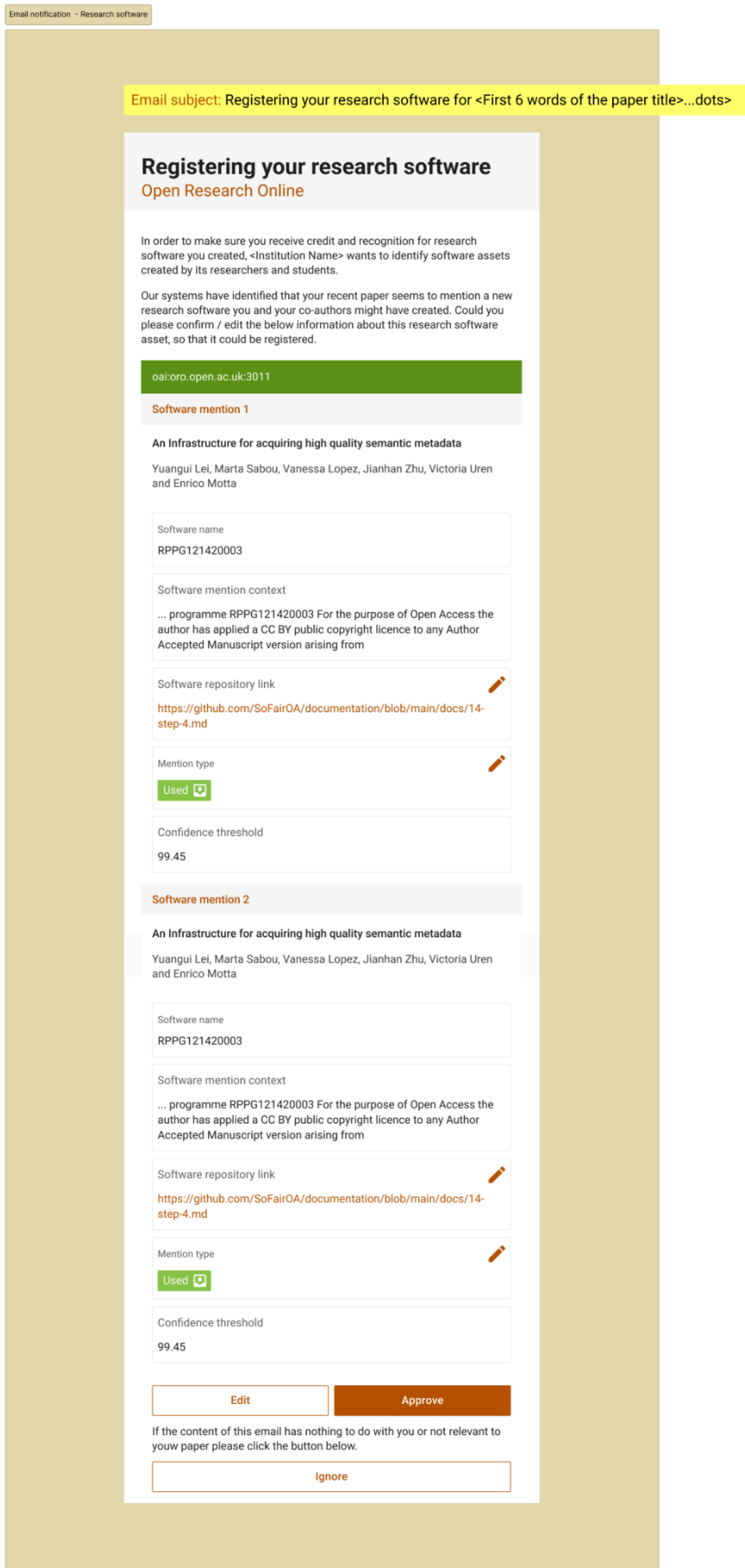

*Figure 3 Sample email for Software Mention validation*

actor: Refers to the entity that initiated the validation process. In this context, it is the repository manager who requested the software mention validation.

object: Contains the software mention validation request, effectively establishing the link between the paper and the software mention.

id: Represents the identifier of the record in the repository where the mention was discovered

sorg:citation: A description of the software following the CodeMeta standard.

mentionConfidence, mentionType, mentionContext: Specifies the type of software usage, the confidence of the detection and the context.

origin: Details the repository endpoint from which the mention originated.

target: Specifies where the results of the activity should be directed.

```json
{
  "@context": [
    "https://www.w3.org/ns/activitystreams",
    "https://purl.org/coar/notify"
  ],
  "actor": {
    "id": "mailto:library@repo.com",
    "name": "Repository manager",
    "type": "Person"
  },
  "id": "urn:uuid:0370c0fb-bb78-4a9b-87f5-bed307a509dd",
  "object": {
    "id": "https://research-organisation.org/repository/record/201203/421/",
    "ietf:cite-as": "https://doi.org/10.5555/12345680",
    "sorg:citation": {
      "@context": "https://doi.org/10.5063/schema/codemeta-2.0",
      "type": "SoftwareSourceCode",
      "name": "SoFAIR",
      "referencePublication": "https://doi.org/10.1016/j.procs.2012.04.202"
    },
    "mentionConfidence": 99,
    "mentionType": "used",
    "mentionContext": "In this paper, we present the software X vY (http://sw/link)"
  },
  "origin": {
    "id": "https://research-organisation.org/repository",
    "inbox": "https://research-organisation.org/inbox/",
    "type": "Service"
  },
  "target": {
    "id": "https://review-service.com/system",
    "inbox": "https://review-service.com/inbox/",
    "type": "Service"
  },
  "type": [
    "Offer",
    "coar-notify:RelationshipAction"
  ]
}
```

*Table 1 COAR Notify ReviewAction payload*

actor: Refers to the entity that is expressing the relationship. In this case CORE.

object: Contains the software mention validation request, effectively establishing the link between the paper and the software mention.

id: Represents the identifier of the record in the repository where the mention was discovered inReplyTo: Represent the identifier of the Relationship offer.

sorg:citation: A description of the software following the CodeMeta standard.

mentionConfidence, mentionType, mentionContext: Specifies the type of software usage, the confidence of the detection and the context.

origin: Details the repository endpoint from which the mention originated.

target: Specifies where the results of the activity should be directed.

```json
{
    "@context": [
        "https://www.w3.org/ns/activitystreams",
        "https://coar-notify.net"
    ],
    "actor": {
        "id": "https://review-service.com/system",
        "name": "CORE",
        "type": "Service"
    },
    "context": {
        "id": "https://research-organisation.org/repository/preprint/201203/421/"
    },
    "id": "urn:uuid:94ecae35-dcfd-4182-8550-22c7164fe23f",
    "inReplyTo": "urn:uuid:0370c0fb-bb78-4a9b-87f5-bed307a509dd",
    "object": {
        "id": "https://research-organisation.org/repository/record/201203/421/",
        "ietf:cite-as": "https://doi.org/10.5555/12345680",
        "sorg:citation": {
            "@context": "https://doi.org/10.5063/schema/codemeta-2.0",
            "type": "SoftwareSourceCode",
            "name": "SoFAIR",
            "referencePublication": "https://doi.org/10.1016/j.procs.2012.04.202"
        },
        "mentionConfidence": 99,
        "mentionType": "used",
        "mentionContext": "In this paper, we present the software X vY (http://sw/link)"
    },
    "origin": {
        "id": "https://review-service.com/system",
        "inbox": "https://review-service.com/inbox/",
        "type": "Service"
    },
    "target": {
        "id": "https://research-organisation.org/repository",
        "inbox": "https://research-organisation.org/inbox/",
        "type": "Service"
    },
    "type": [
        "Announce",
        "coar-notify:RelationshipAction"
    ]
}
```

*Table 2 COAR Notify Announce payload*